\documentclass[a4paper,11pt]{article}
\usepackage{pos}
\usepackage{mathrsfs}
\usepackage{amsmath}
\usepackage{subcaption}
\usepackage{amsfonts}
\usepackage{booktabs}
\usepackage{siunitx}
\usepackage{array}
\newcolumntype{M}[1]{>{\centering\arraybackslash}m{#1}}
\usepackage{verbatim}
\usepackage{float}

\def\bea{\begin{eqnarray}}
\def\eea{\end{eqnarray}}
\def\be{\begin{equation}}
\def\ee{\end{equation}}

\newcommand{\fig}{Fig.~}

\newcommand{\eq}{Eq.~}

\newcommand{\se}{Sec.~}
\newcommand{\ses}{Secs.~}
\newcommand{\re}{Ref.~}

\title{A stabilizing kernel for complex Langevin simulations of real-time gauge theories}
\ShortTitle{A stabilizing kernel for complex Langevin simulations of real-time gauge theories}

\author[a]{Kirill Boguslavski}
\author*[a]{Paul Hotzy}
\author[a]{David I.\ M\"uller}

\affiliation[a]{Institute for Theoretical Physics, TU Wien, \\ 
  Wiedner Hauptstraße 6-8, A-1040 Vienna, Austria}

\emailAdd{kirill.boguslavski@tuwien.ac.at}
\emailAdd{paul.hotzy@tuwien.ac.at}
\emailAdd{dmueller@hep.itp.tuwien.ac.at}

\abstract{The complex Langevin (CL) method is a promising approach to overcome the sign problem, which emerges in real-time formulations of quantum field theories. Over the past decade, stabilization techniques for CL have been developed with important applications in finite density QCD. However, they are insufficient for SU($N_c$) gauge theories on a Schwinger-Keldysh time contour that is required for a real-time formulation. In these proceedings we revise the discretization of the real-time CL equations and introduce a novel anisotropic kernel that enables CL simulations on discretized time contours. Applying it to SU(2) Yang-Mills theory in 3+1 dimensions, we obtain unprecedentedly stable results that may allow us to calculate real-time observables from first principles.}

\FullConference{
  The 39th International Symposium on Lattice Field Theory \\
  August 8 – 13, 2022 \\
  Bonn, Germany
}

\begin{document}
\maketitle

\section{Introduction}
    In recent years the complex Langevin (CL) method has shown promising potential for the calculation of the equation of state of Quantum Chromodynamics (QCD) \cite{Sexty:2019vqx, Attanasio:2022mjd}.
    More generally, it is a powerful approach for the calculation of expectation values in systems suffering from the sign problem, where standard Monte-Carlo integration methods are not applicable. 
    Such a situation occurs for real-time simulations of quantum field theories, where the CL method has been used \cite{Berges:2006xc, Berges:2007nr,Alvestad:2021hsi,Kades:2021hir}. Here we conduct real-time simulations of non-Abelian gauge theories on the Schwinger-Keldysh contour using the CL method.

    In CL the degrees of freedom are complexified to formulate the complex Langevin equation
    \begin{align}
        \mathrm{Re} \, \dot{ z }(\theta) &= \mathrm{Re}\,  K(z(\theta)) + \eta(\theta), \qquad &&
        \textit{Drift term:}\quad K(z)=i\frac{dS}{dz}, ~ z \in \mathcal{M}_C = \mathbb{C},\\
        \mathrm{Im} \, \dot{ z }(\theta) &= \mathrm{Im}\,  K(z(\theta)), \qquad 
        && \textit{Noise term:} \quad \langle \eta(\theta) \rangle = 0,~ \langle \eta(\theta) \eta(\theta^\prime) \rangle = 2 \delta(\theta-\theta^\prime),
    \end{align}
    for some stochastic process $z(\theta)$. Under certain assumptions, the stochastic process described by the CL equation converges to the stationary solution of the complex Fokker-Plank equation \cite{Nagata:2016vkn}. This enables the calculation of expectation values by sampling at large Langevin times $\theta$ 
    \begin{align} \label{eq:sample}
        \langle \mathscr{O} \rangle = \frac{1}{Z} \int d x\, \mathscr{O}(x) \exp \left[ i S(x) \right] \approx \lim\limits_{\theta_0\rightarrow\infty} \frac{1}{T}\int_{\theta_0}^{\theta_0+T}  d \theta\, \mathscr{O}[z(\theta)].
    \end{align}
    However, CL suffers from two types of instabilities. Runaway instabilities can be removed by the introduction of adaptive step sizes \cite{Aarts:2009dg}. Convergence to the wrong stationary solution is not yet resolved in general but could by alleviated by modern stabilization techniques such as gauge cooling \cite{Seiler:2012wz} and dynamical stabilization \cite{Attanasio:2018rtq} or by using appropriately designed kernels \cite{Damgaard:1987rr}.

    In this work and our upcoming publication \cite{boguslavskiHotzyMueller:2022} we present our recent advancements in stabilizing CL in the context of real-time SU($N_c$) Yang-Mills simulations in 3+1 dimensions using the Schwinger-Keldysh formalism.
    After introducing the lattice CL method in Sec.~\ref{sec:lattice}, we review existing stabilization techniques in Sec.~\ref{sec:stabilization}. 
    Previous studies \cite{Berges:2006xc} of CL applied to real-time Yang-Mills theory without any stabilization methods suffered from problems with wrong convergence. In \cite{Berges:2007nr} it is shown that gauge fixing helps convergence at large inverse coupling. Nevertheless, we find that even the application of more recent stabilization techniques yields incorrect results in many cases. We therefore develop a novel anisotropic kernel in Sec.~\ref{sec:new_method} which may provide a systematic approach to avoid instabilities and enable convergence to correct results.
    We demonstrate the effectiveness of our new method in \ses\ref{sec:results} and \ref{sec:monitoring} by comparing to previous results. We conclude in \se\ref{sec:conclusion}.

\section{Complex Langevin method for real-time Yang-Mills theory}
\begin{figure} [t]
        \centering
        \begin{subfigure}[t]{.45\textwidth}
            \centering\includegraphics[width=1\linewidth]{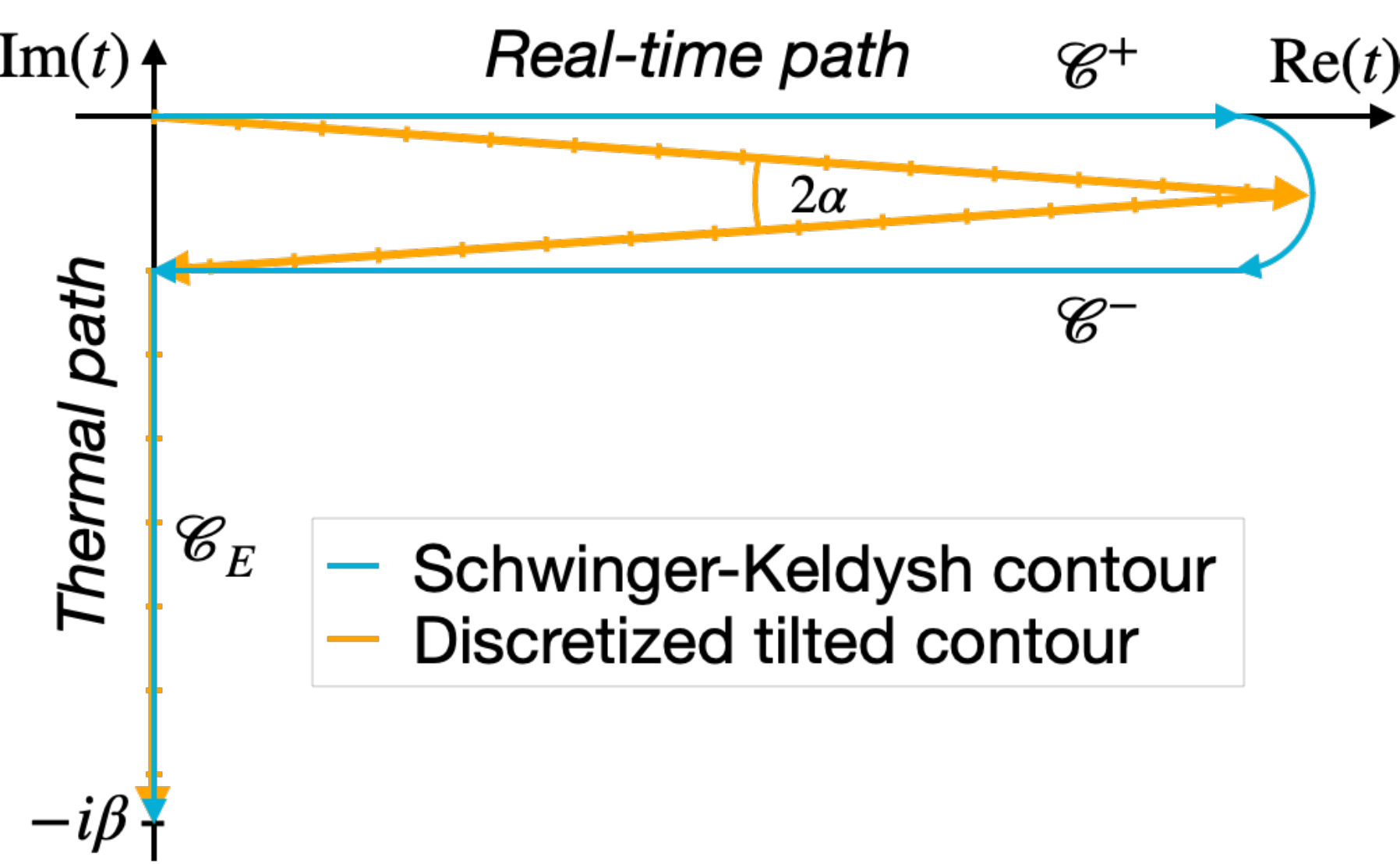}
            \caption{\label{fig:disc_vs_cont_contour}}
        \end{subfigure}
        \quad
        \begin{subfigure}[t]{.45\textwidth}
            \centering\includegraphics[width=1.\linewidth]{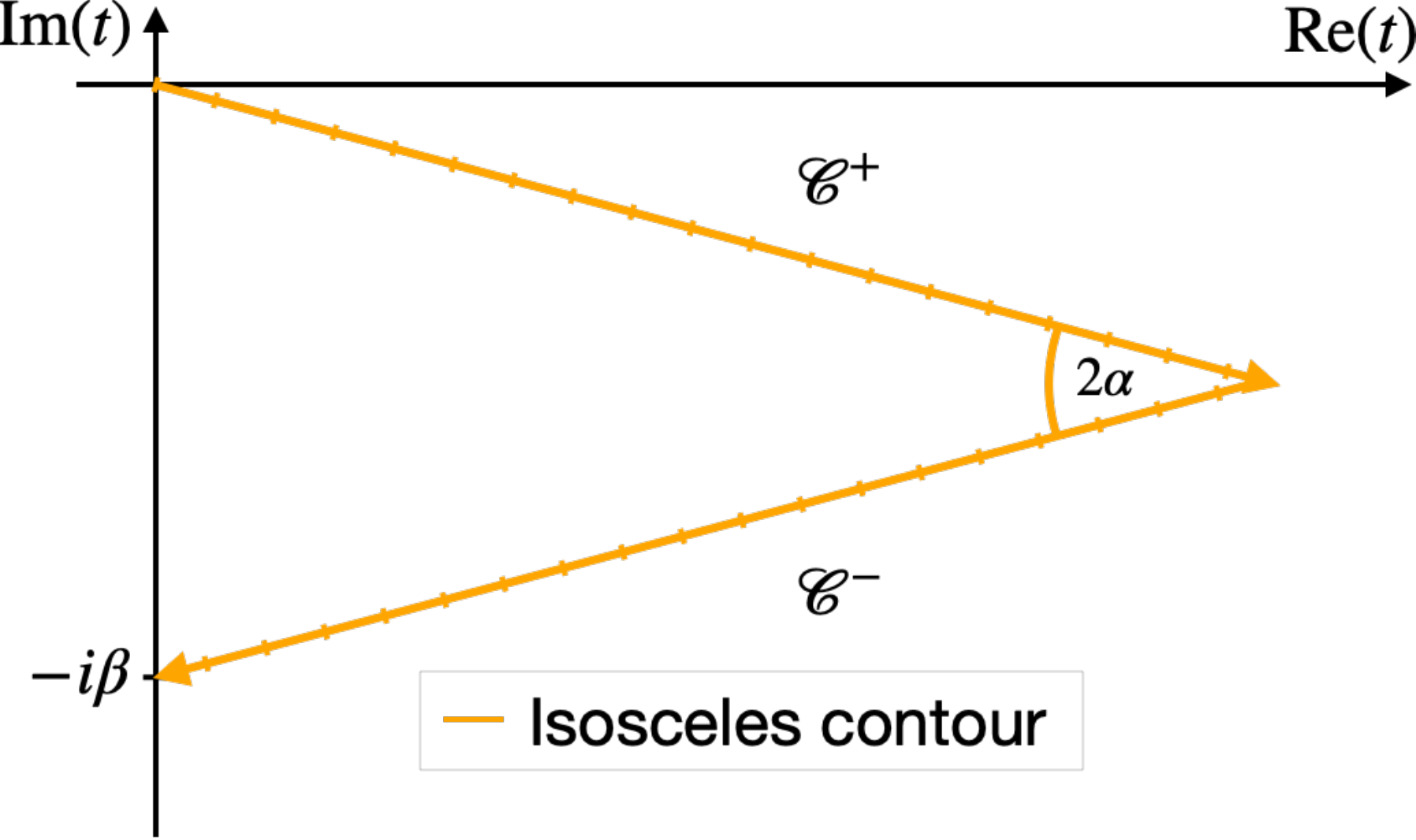}
            \caption{\label{fig:isosceles_contour}}
        \end{subfigure}
        \caption{(a) Schwinger-Keldysh time contour and discretized tilted time-contour. (b) Isosceles time contour used in \re\cite{Berges:2006xc} with tilt angle $\alpha$.
    \label{fig:contour}}
    \end{figure}

    The CL equation for gauge fields in the continuum reads
    \begin{align} \label{eq:cle}
        \frac{\partial A^a_\mu(\theta, x)}{\partial \theta} &= - \frac{\delta S_\mathrm{YM}}{\delta A^a_\mu(\theta, x)} + \eta^a_\mu(\theta, x), \qquad S_\mathrm{YM} = - \frac{1}{4} \int_{\mathscr C} d^4x F^{\mu \nu}_aF_{\mu \nu}^a,
    \end{align}
    where $F_{\mu \nu}^a$ denotes the field strength tensor and Lorentz indices $\mu,\nu = 0,1,2,3$ and color indices $a = 1,\dots,N_c^2-1$ are summed over implicitly. In general, this equation is not unique but is a representative of an equivalence class of evolution equations that converge to the same stationary solution \cite{Damgaard:1987rr}.
    We will utilize this so-called kernel freedom in Sec.\ \ref{sec:new_method} in order to stabilize CL simulations. For the CL method the gauge fields $A^a_\mu$ are complexified and therefore form the $\mathfrak{sl}(2, \mathbb{C})$ Lie algebra. The degrees of freedom of the gauge fields $A^a_\mu$ are taken into account by the Gaussian distributed noise term
    \begin{align} \label{eq:noise}
        \langle \eta^a_\mu(\theta, x) \rangle &= 0, \quad \langle \eta^a_\mu(\theta, x) \eta^b_\nu(\theta', y) \rangle = 2 \delta(\theta - \theta') \delta^{(d)}(x-y) \delta^{ab} \delta_{\mu\nu}.
    \end{align}
    The complex contour path $\mathscr{C}$ which is integrated over in \eq\eqref{eq:cle} denotes the Schwinger-Keldysh contour and is visualized as the blue curve shown in Fig.\ \ref{fig:disc_vs_cont_contour}. The Schwinger-Keldysch formalism allows us to calculate expectation values via
    \begin{align}
        \langle \mathscr{O}[A] \rangle = \frac{1}{Z}\int \mathcal{D}A_E \, e^{-S_E[A_E]} \int \mathcal{D}A_+\, \mathcal{D}A_-\, e^{i S[A_+,A_-]} \, \mathscr{O}(A)
    \end{align}
    where $A_+$, $A_-$ denote the gauge fields on the forward and backward real-time paths ($\mathscr{C}^+$, $\mathscr{C}^-$) respectively while $A_E$ is defined on the Euclidean (purely imaginary) part of the contour $\mathscr{C}_E$. The gauge fields satisfy periodic boundary conditions
    \begin{align}
        A^a_\mu (t=0) = A^a_\mu (t=-i\beta).
    \end{align}

\subsection{CL simulations on the lattice} \label{sec:lattice}

Following \cite{Berges:2006xc} we discretize the gauge field by introducing link and plaquette variables
     \begin{align}
        U_{x,\mu} &= \exp(iga_\mu A^a_\mu(x+\hat{\mu} / 2) t^a) \, \,  \in \, \, \mathrm{SL}(N_c,\mathbb{C}), \\
        U_{x,\mu \nu}(x) &= U_{x, \mu}U_{x+\mu, \nu}U_{x+\nu,\mu}^{-1} U_{x, \nu}^{-1},
    \end{align}
    on an $N_t \times N_s^3$ lattice and use the Wilson action
     \begin{align}
        S[U] = \frac{1}{2N_c} \sum_{x, \mu, \nu} \beta_{\mu\nu} \mathrm{Tr}\left[U_{x,\mu\nu} - 1\right],
    \end{align}
    where the coupling constants are denoted by $\beta_0=\beta_{0i}=\beta_{i0}=\frac{2N_c}{g^2}\frac{a_i}{a_0}$ and $\beta_s=\beta_{ij}=\beta_{ji}=\frac{2N_c}{g^2}\frac{a_0}{a_i}$ and we assume spatial lattice spacings $a_i=a_s$ for $i=1,2,3$. 

    A commonly used discretization of the CL equation corresponds to the update scheme
    \begin{align} \label{eq:berges_cl}
        U_{x, \mu}(\theta+\epsilon) &= \exp \left( i t^a \left[ 
         -   \epsilon K^a_{x,\mu}(\theta) + \sqrt{ \epsilon} \eta^a_{x,\mu}(\theta)
         \right] \right) U_{x, \mu}(\theta),
         \end{align}
    where the drift term is given by $K^a_{x,\mu} =  \left[ \frac{\delta S}{\delta  A^a_{x,\mu}}\right]_\mathrm{latt}$ defined via the variation of the lattice action
    \begin{align}
        \delta S = \sum_{x} \bigg[ \frac{\delta S}{\delta A^a_{x,\mu}}\bigg]_\mathrm{latt} \delta  A^a_{x,\mu}.
    \end{align}
    As was commented in \cite{Matsumoto:2022ccq}, the discretized path integral needs to be regularized because it is not analytic with respect to the lattice spacing along the time contour. This subtle non-analyticity is resolved by tilting the real-time part of the contour as depicted in Fig.\ \ref{fig:disc_vs_cont_contour}.
    
    In CL we calculate expectation values of observables by averaging uncorrelated values at sufficiently late Langevin times. To guarantee uncorrelated sampling of a particular observable $\mathscr{O}[A]$, we compute the auto-correlation function $R_{\mathscr{O}}$ and auto-correlation time $T_{\mathscr{O}}$ 
    \begin{align}
        R_{\mathscr{O}}(\tau) &= 
        \frac{
            \langle 
            \left( \mathscr{O}_{\theta}- \langle \mathscr{O}_{\theta} \rangle \right)  \left( \mathscr{O}_{\theta + \tau}- \langle \mathscr{O}_{\theta + \tau} \rangle \right) \rangle
        }
        {\sigma_{\theta} \sigma_{\theta + \tau}} \approx \exp\left(-\tau  / T_{\mathscr{O}}\right),
    \end{align}
    where $\sigma_{\theta}$ denotes the standard deviation of the observable at time $\theta$.
    
    \subsection{Stabilization techniques} \label{sec:stabilization}
    
    Complex Langevin simulations are inherently unstable. Several methods to mitigate these instabilies were introduced in recent years. 
    We adapt some of them to the update steps of \eq\eqref{eq:berges_cl}.

\paragraph{Adaptive step size (AS)}
    We adaptively change the step size $\epsilon$ with respect to the maximum drift term relative to a sufficiently small upper bound $B$ 
    \begin{align} \label{eq:as}
        \epsilon \mapsto \tilde \epsilon = 
        \epsilon \min\left(1, \frac{B}{\max\limits_{x,\mu, a} \vert K_{x,\mu}^a\vert} \right).
    \end{align}
    This method was initially introduced in \cite{Aarts:2009dg}.

    Adaptive step sizes can remove runaway instabilities otherwise encountered due to large drift terms pointing towards the bulk of the complex manifold.

\paragraph{Gauge cooling (GC)}
    We further adopt the gauge cooling procedure introduced in \cite{Seiler:2012wz} and developed further in \cite{Aarts:2013uxa} to alleviate instabilities of our CL simulations.
    This method is exploiting gauge freedom by minimizing a gauge dependent functional $F[U]$ which measures the non-unitarity of the configuration. Empirically, it has been shown that this leads to more stable simulations.

    The minimization process is done by gauge transforming the link field configuration
    \begin{align} \label{eq:gc}
        U_{x,\mu} \; \mapsto \; &U_{x, \mu}^V = V_{x,\mu} U_{x, \mu} V_{x+\mu,\mu}^{-1}, \qquad F[U] \geq F[U^V],
    \end{align}
    where the gauge transformation is determined by a gradient descent scheme. 
    It has been argued in \cite{Nagata:2015uga} that gauge cooling does not bias the results of CL for gauge invariant observables. Several different functionals have been used so far in the literature. We use a version of the so called unitarity norm
            \begin{align}
        \label{eq:unorm}
        F[U] = \sum\limits_{x,\mu} \mathrm{Tr}\left[(U_{x, \mu} U_{x, \mu}^\dagger - 1)^2\right],
    \end{align}
    which differs from the original formulation by the inclusion of the square. We find that this speeds up the minimization process.

\paragraph{Dynamical stabilization (DS)}
    Lastly, we test dynamical stabilization \cite{Attanasio:2018rtq} 
    which penalizes the imaginary part of the drift term if the configuration is locally not unitary. We substitute the drift term
    \begin{align}
        &K_{x,\mu}^a \mapsto \tilde{K}_{x,\mu}^a = K_{x,\mu}^a + i \alpha_\mathrm{DS} M_{x}^a, \\
        &M_{x}^a = b_{x}^a \left( \sum_{c} b_{x}^c b_{x}^c \right), \quad 
        b_{x}^a = \sum_{\mu} \mathrm{Tr}[t^a U_{x,\mu} U_{x,\mu}].
    \end{align}
 The force parameter $\alpha_\mathrm{DS}$ is tuned such that observables become approximately independent of it. This method has resulted in advancements in finite density equation of state calculations in QCD  \cite{Attanasio:2022mjd}. However, dynamical stabilization is not rigorously justified yet and we will comment on its applicability to real-time simulations of gauge theories in Sec.~\ref{sec:results}.
    
\section{Progress on real-time Yang-Mills simulations} \label{sec:progress}
In this section we first resolve some ambiguities in the discretized CL update step of Eq.~\eqref{eq:berges_cl} and then introduce a new method, namely an anisotropic kernel, that is able to avoid previous convergence problems \cite{boguslavskiHotzyMueller:2022}. We test our method in Sec.~\ref{sec:results} by comparing the results of one-point functions to simulations on the stable Euclidean time contour and in \se\ref{sec:monitoring} to validation observables including the unitarity norm and Dyson-Schwinger equations. 
Following \cite{Berges:2006xc} we conduct these simulations on a 3+1 dimensional lattice where we neglect the Euclidean part of the tilted time-contour such that the imaginary parts of $\mathscr{C}^+$ and $\mathscr{C}^-$ span from $t=0$ to $t=-i\beta$ (isosceles contour) as shown in Fig.~\ref{fig:isosceles_contour}. We have checked that our method is also applicable to the discretized Schwinger-Keldysh contour in \fig\ref{fig:disc_vs_cont_contour}.
If not stated otherwise we use the SU(2) gauge group, the inverse temperature $\beta = 4.0$, the coupling constant $g=1.0$, and a lattice with $N_s=4$ and $N_t=16$.

\subsection{Discretization of the time contour and introduction of an anisotropic kernel} \label{sec:new_method}

The commonly used lattice discretizion of the CL equation \eqref{eq:berges_cl} in combination with the Schwinger-Keldysh contour is ambiguous due to the complex nature of the time contour. More specifically, it is not obvious how complex arguments in the Dirac distribution in Eq.~\eqref{eq:noise} should be treated. In order to resolve this we parameterize the contour by its arc length $\lambda$    and replace the complex-valued time $t$ with the real-valued parameter $\lambda$ in the noise correlator. Upon discretization of the contour (see Fig.~\ref{fig:disc_vs_cont_contour}) and Eqs.~\eqref{eq:cle} with $\lambda$-spacing $a_\lambda$, we obtain the following update equations
        \begin{align} 
        \label{eq:link_update}
         U_{x, \lambda}(\theta+\epsilon) &= \exp \left( i t^a \left[ 
         -   \epsilon\, \frac{a_\lambda}{a_s} \left[ \frac{\delta S}{\delta  A^a_{x, t}}\right]_\mathrm{latt}(\theta) + \sqrt{ \epsilon} \sqrt{\frac{a_\lambda}{a_s}} \eta^a_{x, \lambda}(\theta)
         \right] \right) U_{x, \lambda}(\theta), \\
        U_{x, i}(\theta+\epsilon) &= \exp \left( i t^a \left[ 
         -  \epsilon\, \frac{a_s}{\bar a_\lambda}  \left[ \frac{\delta S}{\delta  A^a_{x, i}}\right]_\mathrm{latt}(\theta) +  \sqrt{ \epsilon} \sqrt{\frac{a_s}{\bar a_\lambda}} \eta^a_{x, i}(\theta)
         \right] \right) U_{x, i}(\theta)\,.
         \label{eq:link_update_spat}
    \end{align}
    The averaged lattice spacing $\bar{a}_\lambda = \frac{a_\lambda + a_{\lambda-1}}{2}$ recovers time-reversal symmetry for the spatial link update. Since $\lambda$ describes the arc length, we use $a_\lambda = |a_t|$ in the discretized setting. The temporal links $U_{x,0}$ of Eq.~\eqref{eq:berges_cl} are replaced by links along the contour $U_{x,\lambda}$ and we note that we recover the original update equations by setting $a_\lambda = a_s$ \cite{boguslavskiHotzyMueller:2022}.
    
    In addition, we exploit the kernel freedom of the CL equation (see e.g.~chapter 4 of \cite{Namiki:1993fd}) by introducing a field independent kernel which effectively rescales the Langevin time step for the temporal update by $a_\lambda / a_s$ and the spatial link update by $\bar a_\lambda / a_s$ (see Eqs.~(\ref{eq:link_update}, \ref{eq:link_update_spat}) for comparison). If one ignores the subtle difference between $a_\lambda$ and $\bar a_\lambda$, our kernel corresponds to a simple rescaling of the Langevin step $\epsilon$. We obtain
    \begin{align} \label{eq:new_cle_temp}
         U_{x, \lambda}(\theta+\epsilon) &= \exp \left( i t^a \left[ 
         -  \epsilon \left(\frac{a_\lambda}{a_s}\right)^2 \left[ \frac{\delta S}{\delta  A^a_t}\right]_\mathrm{latt}(\theta) + \sqrt{ \epsilon}\, \frac{a_\lambda}{a_s} \eta^a_{x,\lambda}(\theta)
         \right] \right) U_{x, \lambda} (\theta), \\ \label{eq:new_cle_spat}
        U_{x, i}(\theta+\epsilon) &= \exp \left( i t^a \left[ 
         - \epsilon   \left[ \frac{\delta S}{\delta A^a_i}\right]_\mathrm{latt}(\theta) +  \sqrt{ \epsilon} \, \eta^a_{x,i}(\theta)
         \right] \right) U_{x, i} (\theta),
    \end{align}
    for our kerneled update steps. Comparing these new update equations (\ref{eq:new_cle_temp}, \ref{eq:new_cle_spat}) to the commonly used method in \eq\eqref{eq:berges_cl}, our modification amounts to an anisotropic kernel which rescales only the temporal links. The motivation of the form of this kernel is twofold. We notice that the noise term of the spatial update step in \eq\eqref{eq:link_update_spat} blows up in the temporal continuum limit $a_\lambda\rightarrow 0$ when the Langevin time step $\epsilon$ is held constant. Therefore, we rescale the spatial update step to remove this behavior. Secondly, we observe that the fluctuations of the temporal link fields are small compared to the spatial directions. Hence, this allows us to upscale the Langevin time step for the temporal updates.

    \subsection{Improved stability and convergence of our kerneled CL equation} \label{sec:results}
    
\begin{figure} [t]
        \centering
        \begin{subfigure}[t]{.49\textwidth}
            \centering\includegraphics[width=1\linewidth]{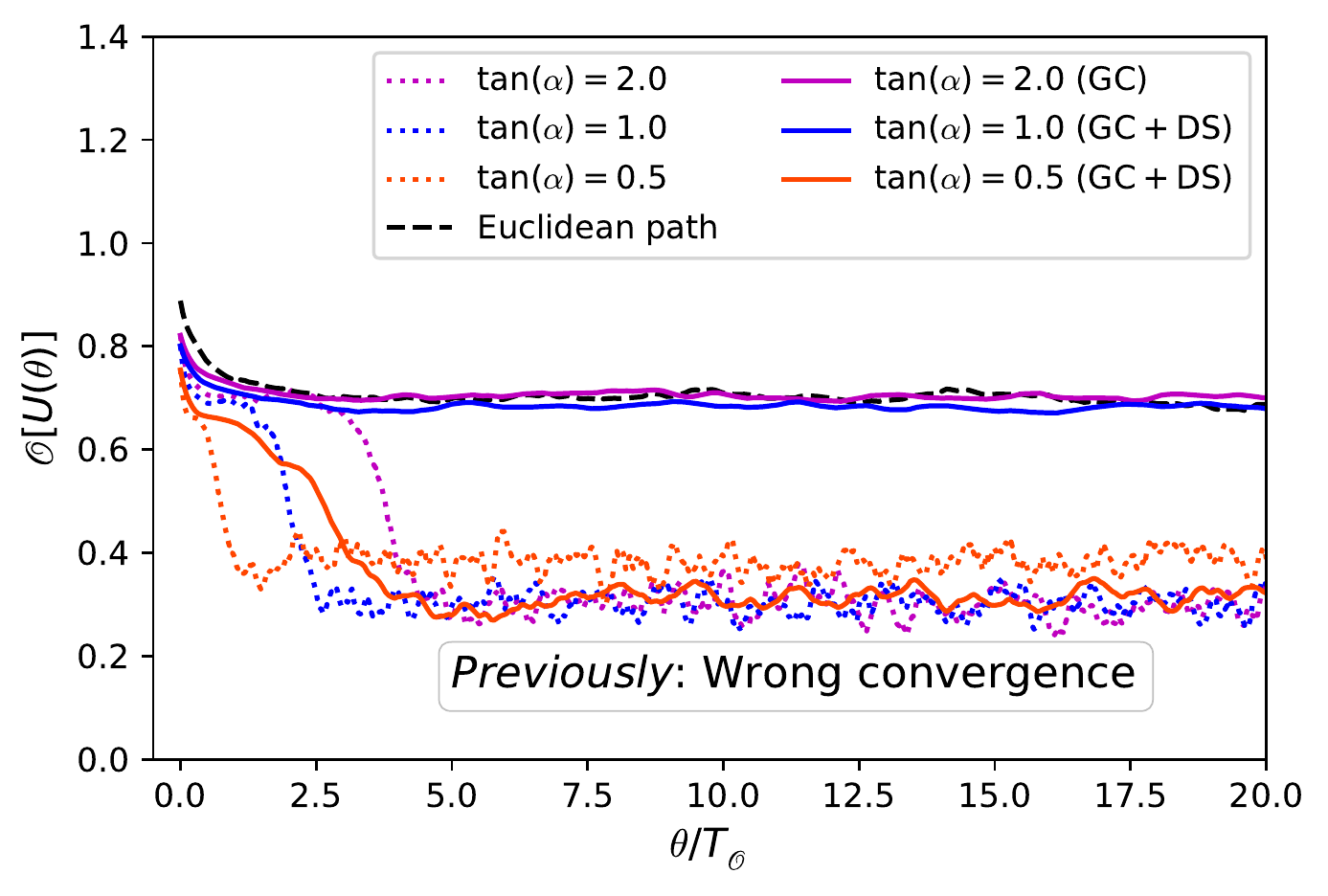}
            \caption{\label{fig:avg_spat_plaq_a}}
        \end{subfigure}
        \begin{subfigure}[t]{.49\textwidth}
            \centering\includegraphics[width=1.\linewidth]{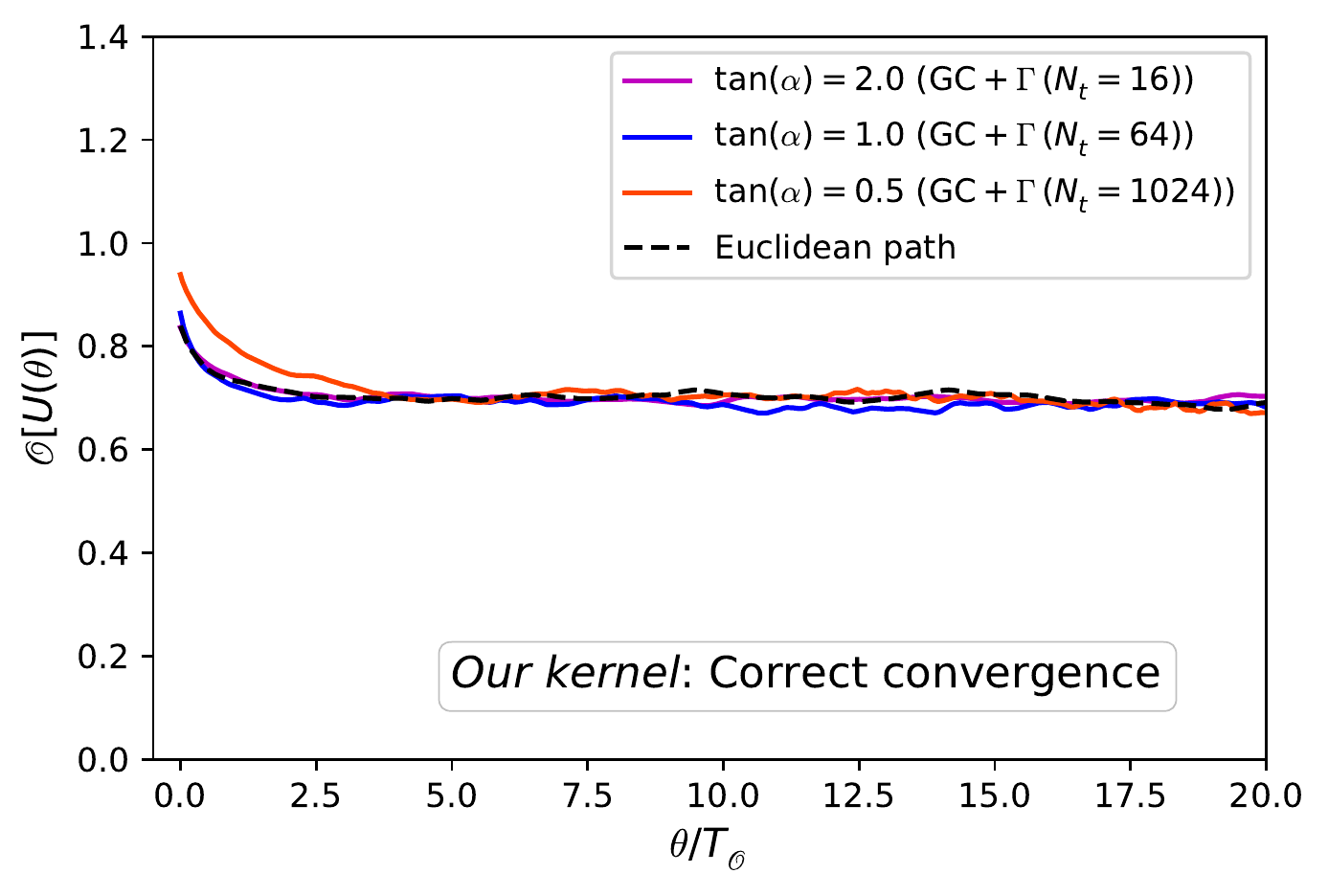}
            \caption{\label{fig:avg_spat_plaq_b}}
        \end{subfigure}
        \caption{(a) Results for the real trace of the average spatial plaquette $\mathscr{O}$ for different tilt angles $\alpha$ of the discretized Schwinger-Keldysch contour without Euclidean path and various stabilization techniques. (b) Results obtained using our anisotropic kernel denoted by $\Gamma(N_t)$. {\em (Both panels)} For results that converge correctly, the Langevin time is rescaled by the autocorrelation time of the stable region. All simulations are initialized by configurations of unit matrices and with the same seed for the random number generator. Simulations are evolved up to $\theta / T_\mathscr{O} = 50$, but we focus on the interval $\theta / T_\mathscr{O} \in [0, 20]$ in order to resolve instabilities towards wrong convergence. \label{fig:avg_spat_plaq}}
    \end{figure}

    We discuss the improvements using the example of the real trace of the average spatial plaquette 
    \begin{align}
        \mathscr{O}[U]=\frac{1}{N_t N_s^3}\sum_x \frac{1}{3} \sum_{i < j} \frac{1}{N_c} \mathrm{Re} \mathrm{Tr} \, U_{x,ij} \, .
    \end{align}
    We first reproduce results of \re\cite{Berges:2006xc} as dotted curves in \fig\ref{fig:avg_spat_plaq_a} for different tilt angles $\alpha$ on the isosceles time contour. In the entire \fig\ref{fig:avg_spat_plaq} we scale the Langevin time with the auto-correlation time of the plotted observable. This is important as we only want to sample uncorrelated data. Additionally, we use moving averages in all of our figures to smoothen the curves. The figure shows that without any additional stabilization, CL converges to a wrong result. The values for the tilted contour should be consistent with the simulation of the Euclidean path because of the time translation invariance of thermal systems. We note that the Euclidean (purely imaginary) path can be simulated without instabilities due to the absence of the sign problem.

    We also observe in \fig\ref{fig:avg_spat_plaq_a} that the gauge cooling procedure stabilizes contours with sufficiently large tilt angles $\alpha$. However, when applied to smaller tilt angles $\tan(\alpha)=1, 0.5$, GC only mitigates instabilities and the process converges to the same (wrong) results as before. Similarly, dynamical stabilization can be applied for large tilts but introduces a small bias for stabilized results due the penalty term.
    For the small tilt angle $\tan(\alpha)=0.5$ as shown in Fig.\ \ref{fig:avg_spat_plaq_a} the penalty term negatively impacts the dynamics due to the rapid increase of the unitarity norm and leads to wrong results. In this case we found no interval where the observable is insensitive to the force parameter $\alpha_{\mathrm{DS}}$.

    \begin{table}[htb!]
        \centering
        \caption{Expectation values of the average spatial plaquette using CL on an isosceles contour with tilt angle $\alpha$. We indicate different stabilization techniques and number of temporal lattice sites used in the simulation. The values are calculated using one simulation run with the same seed for the random number generator. \label{tab:avg_spat_plaq}}
        \begin{tabular}{M{0.2\textwidth} | M{0.3\textwidth} M{0.1\textwidth} M{0.2\textwidth}} \toprule
        {$\tan(\alpha)$} & {Stabilization techniques} & {$N_t$} & {$\langle\mathscr{O}\rangle$} \\ \midrule
        {Euclidean} & {None} & {16} & {$0.704 \pm 0.002$} \\ \midrule
        {2.0} & {AS, GC} & {16} & {$0.701 \pm 0.002$} \\ 
        {1.0} & {AS, GC, DS} & {16} & {$0.678 \pm 0.002$} \\
        {0.5} & {AS, GC, DS} & {16} & {$0.318 \pm 0.007$} \\ \midrule
        {2.0} & {AS, GC, $\Gamma$} & {16} & {$0.701 \pm 0.003$} \\ 
        {1.0} & {AS, GC, $\Gamma$} & {64} & {$0.703 \pm 0.003$} \\ 
        {0.5} & {AS, GC, $\Gamma$} & {1024} & {$0.709 \pm 0.004$} \\ \bottomrule
        \end{tabular}
    \end{table}
    
    In Fig.\ \ref{fig:avg_spat_plaq_b} we show results obtained with our novel anisotropic kernel. Increasing the number of lattice sites $N_t$ along the time contour in conjunction with our kernel successfully stabilizes smaller tilt angles without introducing a bias. For these results we apply the gauge cooling procedure after each CL step but do not use dynamical stabilization. We emphasize that merely increasing $N_t$ without using our kernel leads to similar behaviour as for unstabilized simulations.
    Empirically, we find that the range of stability in $\theta$ grows faster than the auto-correlation time with the number of temporal lattice sites $N_t$. This enables us to postulate a systematic approach to mitigate the instabilities by carrying out a partial (temporal) continuum limit for smaller tilt angles. 
    
    Table \ref{tab:avg_spat_plaq} lists the expectation values of the real trace of the average spatial plaquette for various simulations. The data shows that dynamical stabilization introduces a bias to the result and even breaks down for small tilt angles. Our anisotropic kernel yields values in good agreement with the Euclidean result for all tested tilt angles.

    \subsection{Validating observables: Dyson-Schwinger equations and unitarity norm}
    \label{sec:monitoring}

\begin{figure} [t]
        \centering
        \begin{subfigure}[t]{.49\textwidth}
            \centering\includegraphics[width=1\linewidth]{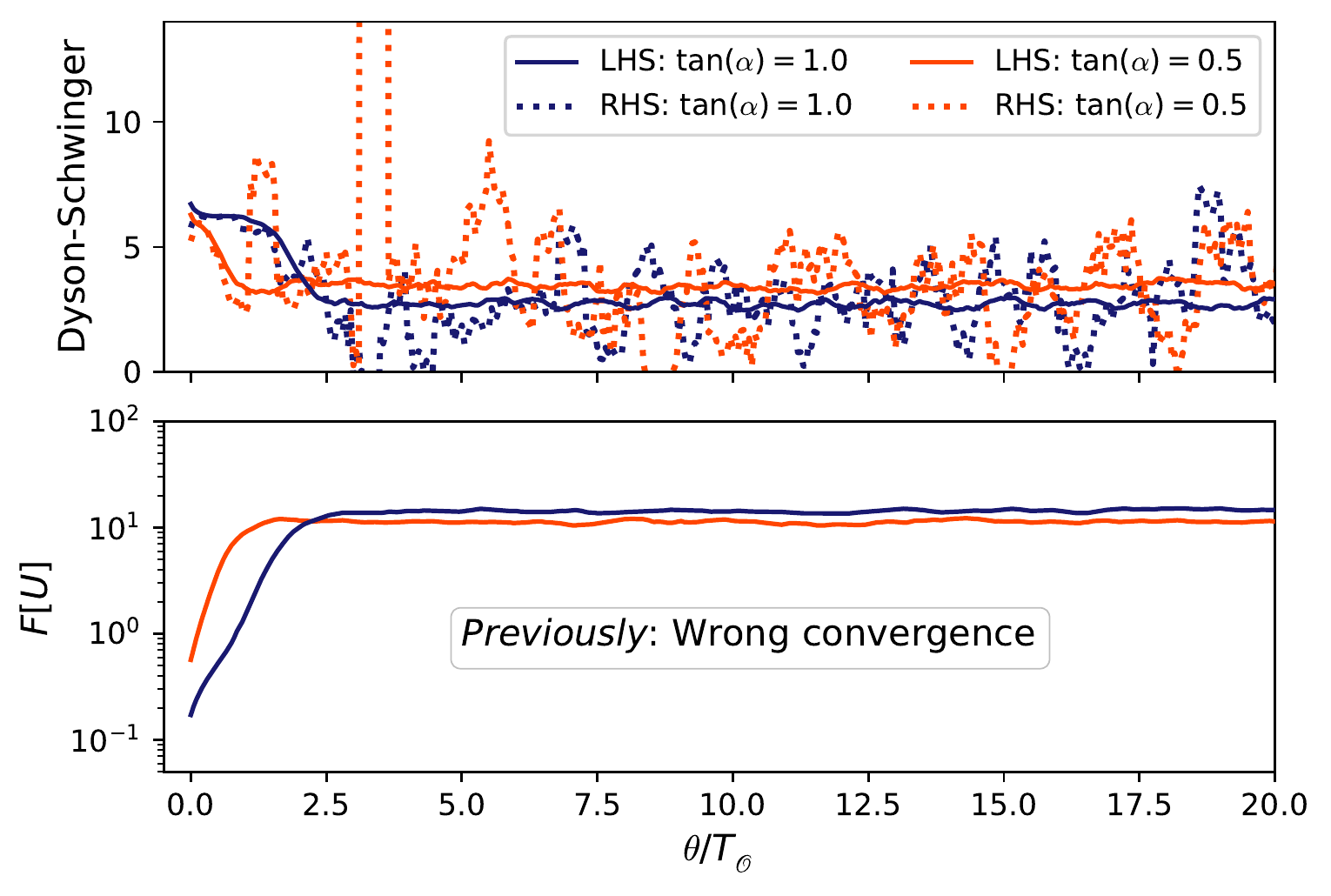}
            \caption{\label{fig:dse_spat_plaq_a}}
        \end{subfigure}
        \begin{subfigure}[t]{.49\textwidth}
            \centering\includegraphics[width=1.\linewidth]{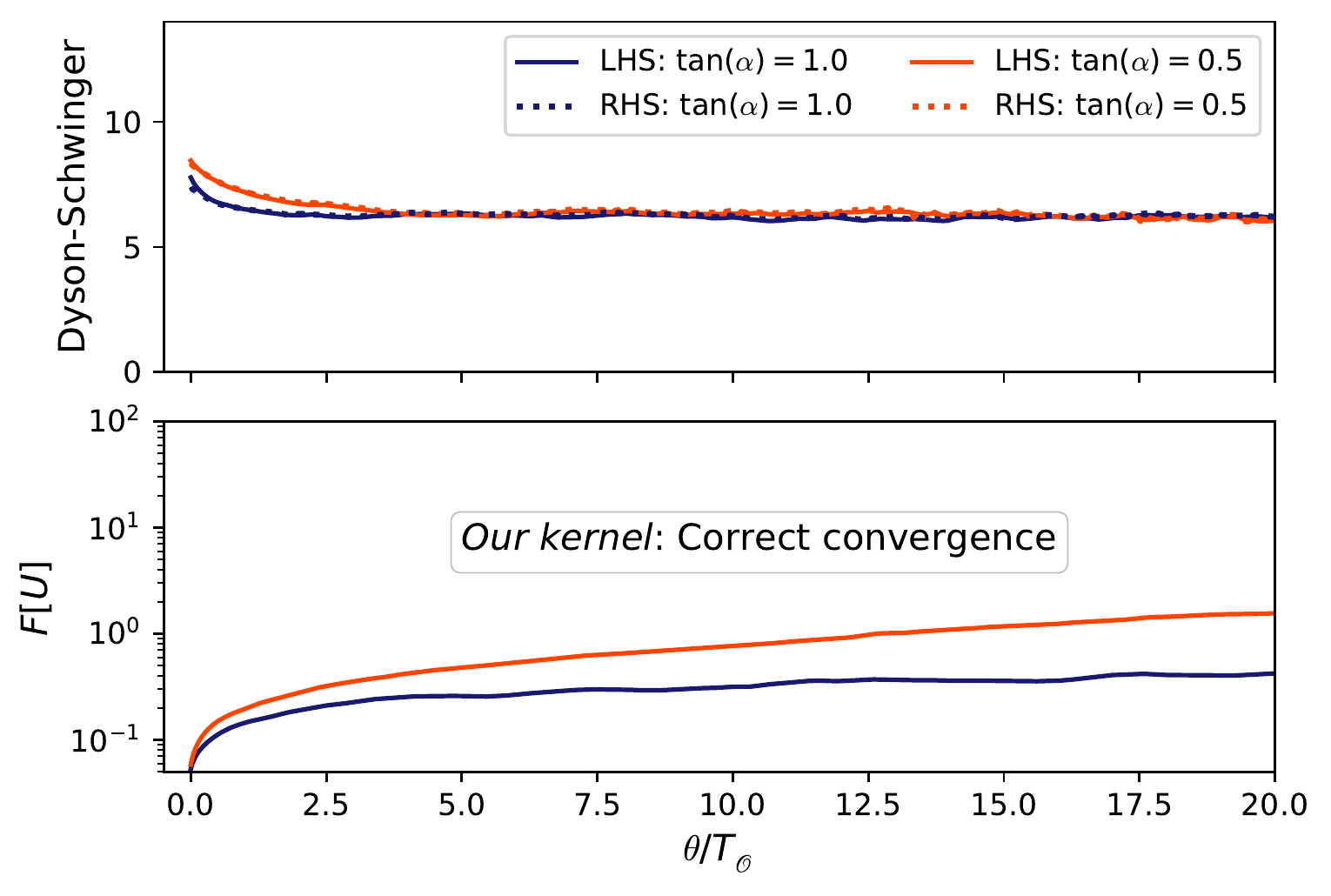}
            \caption{\label{fig:dse_spat_plaq_b}}
        \end{subfigure}
        \caption{(a) Our results for the left- and right-hand side of the Dyson-Schwinger equation of the average spatial plaquette (top) and the unitarity norm (bottom) only using an adaptive step size. 
        (b) The same with our anisotropic kernel and gauge cooling. We use the same discretization as in Fig.~\ref{fig:avg_spat_plaq}. \label{fig:dse_spat_plaq}}
    \end{figure}

    In addition to our comparison to Euclidean simulations, we validate our results using  the Dyson-Schwinger equations for spatial plaquettes
    \begin{align}
        \begin{split}
            \frac{2(N_c^2-1)}{N_c} \sum_{i<j} \left\langle \mathrm{ReTr}(U_{x, ij}) \right\rangle = 
            \frac{i}{2N_c} \sum_{i<j} \sum_{\vert \rho \vert \neq i} \beta_{i \rho} 
            \left\langle \mathrm{ReTr} 
            \left[ 
            ( U_{x,i\rho} + U_{x,i\rho}^{-1} ) U_{x,ij}
            \right] 
            \right\rangle
        \end{split}
    \end{align}
    and the unitarity norm in \eq\eqref{eq:unorm}.
    The top panels of Fig.~\ref{fig:dse_spat_plaq} show a comparison of the results for the left-hand side (LHS) and right-hand side (RHS) of the Dyson-Schwinger equations. Since these equations have to be satisfied identically, we utilize them to benchmark our CL simulation by checking self-consistency of the evolved link configurations.
    Figure \ref{fig:dse_spat_plaq_a} reproduces similar results to \re\cite{Berges:2006xc} without stabilization. The authors of \cite{Berges:2006xc} showed that the equations approximately hold even for wrong convergence results, although the RHS tends to exhibit large fluctuations. 
    On the other hand, Fig.~\ref{fig:dse_spat_plaq_b} shows that our kernel does not introduce any bias and produces stable results even for the RHS.
    
    The bottom panels of Fig.~\ref{fig:dse_spat_plaq} show the unitarity norm of the unstable simulations (a) and simulations using our new method (b). Combining gauge cooling with our kernel leads to a reduction of the unitarity norm by an order of magnitude and a weaker increase over time. As a result, we obtain an enhanced region of correct convergence that we can sample over efficiently.

\section{Conclusion} \label{sec:conclusion}
    
    We have studied the CL method applied to real-time simulations of non-Abelian Yang-Mills theories and introduced a novel approach that led to unprecedentedly stable results on discretized time contours.
    In particular, without additional stabilization, real-time CL simulations suffer from severe instabilities and convergence to wrong results. 
    We have shown that modern stabilization techniques including adaptive step sizes, gauge cooling and dynamical stabilization mitigate these problems but break down at decreasing tilt angles of the discretized time contour. 

    We therefore put forward a novel anisotropic kernel. It effectively rescales the Langevin time step in spatial and temporal directions, which enlarges stable regions of correct convergence for smaller tilt angles at the cost of finer temporal lattice discretizations.  In \cite{Matsumoto:2022ccq} it was argued that the order of the limits of taking first a finer temporal discretization and a subsequently decreasing tilt is important to correctly regularize the discretized path integral of gauge theories. 

    Our kernel thus appears to be tailored to exactly this program. This promising approach may enable us to calculate real-time observables on a continuous Schwinger-Keldysh contour. In our upcoming publication \cite{boguslavskiHotzyMueller:2022}, we introduce our new technique in more detail and investigate the prospect to compute real-time observables directly in our simulations.

\begin{acknowledgments}
  The authors are grateful to D.~Alvestad, J.M.~Pawlowski, D.~Sexty and F.~Ziegler for valuable discussions and very useful comments, and to A.\ Ipp for technical input regarding code development.
  This research was funded by the Austrian Science Fund (FWF) project P~34455-N. Paul Hotzy furthermore expresses his gratitude to the Doktoratskolleg Particles and Interactions (DK-PI,  FWF doctoral program No.~W-1252-N27) which supported his research and thereby this project.
  The computational results presented have been achieved in part using the Vienna Scientific Cluster (VSC).
\end{acknowledgments}

\bibliographystyle{JHEP}
\bibliography{lattice_pos}

\end{document}